\documentclass[twocolumn,floatfix,showpacs,preprintnumbers,superscriptaddress]{revtex4}
\usepackage{amsfonts,amsthm,amssymb,latexsym}
\usepackage{natbib}
\usepackage{dcolumn}
\usepackage{bm}
\usepackage[fleqn]{amsmath}
\usepackage[spanish]{babel}
\usepackage[latin1]{inputenc}
\usepackage[dvips]{graphicx}
\usepackage[centerlast,small]{caption2}

\begin{document}
\title{\textbf{SOBRE LA ECUACI\'ON DE TRANSFERENCIA RADIATIVA RELATIVISTA
ESPECIAL}}
\author{Mauricio F. Duque D.}
\email{mfduqued@unal.edu.co}
\affiliation{Departamento de F\'{\i}sica, Universidad Nacional de Colombia, Bogot\'a,
Colombia}
\author{Leonardo Casta\~neda C.} \altaffiliation[Presente direcci\'on: ]{AIfA, Bonn, Germany}
\email{lcastanedac@unal.edu.co -- leonardo@astro.uni-bonn.de}
\affiliation{Observatorio Astron\'omico Nacional, Universidad Nacional de Colombia,
Bogot\'a, Colombia}
\author{Carlos A. Duque D.}
\email{caduqued@unal.edu.co}
\affiliation{Departamento de Matem\'aticas, Universidad Nacional de Colombia, Bogot\'a,
Colombia}
\affiliation{Departamento de Ingener\'{\i}a Mec\'anica y Mecatr\'onica, Universidad
Nacional de Colombia, \\
Bogot\'a, Colombia}

\begin{abstract}
Con el fin de introducir de una manera clara y directa a los estudiantes de pregrado en f\'{\i}sica y/o astronom\'{\i}a en
el tema de la transferencia radiativa, se realiza una revisi\'on pedag\'ogica explicando la manera de obtener la ecuaci\'on
de transferencia radiativa, sus restricciones y los diferentes tipos de interacci\'on presentes entre la radiaci\'on y la
materia. Debido a que en la literatura encontrada sobre transferencia radiativa la covarianza no es expl\'{\i}citamente
desarrollada, se hace necesario mostrar de manera expl\'{\i}cita los c\'alculos en detalle y discutir sobre los efectos
relativistas especiales.\newline Descriptores: Transferencia Radiativa, Relatividad Especial.\newline

The purpose is to introduce in a clear and direct way the students of undergraduate courses in physics and/or astronomy
to the subject of radiative transfer. A pedagogical revision is made in order to obtain the radiative transfer equation, its
restrictions and the different types of interactions present between the
radiation and the matter. Because  in the classical
literature about radiative transfer the covariance is not fully developed, we show in an explicit
manner detail calculations and then we discuss the relativistic effects.

.\newline Keywords: Radiative Transfer,
Special Relativity.
\end{abstract}

\maketitle

\section{Introducci\'on}
La teor\'{\i}a de la transferencia radiativa fue desarrollada por S. Chandrasekhar, en trabajos desarrollados entre 1944 a
1949, y expuesta en su texto \cite{chandrasekhar01}; en esta teor\'{\i}a, se describe la dispersi\'on y absorci\'on de la
radiaci\'on electromagn\'etica por el medio interestelar \cite{ribicky}. Dicha descripci\'on se puede realizar en la
aproximaci\'on fenomenol\'ogica usando la variable intensidad espec\'{\i}fica \cite{mihalas1} o en la aproximaci\'on
cinem\'atica por medio de la funci\'on de distribuci\'on. En la actualidad el estudio de la transferencia radiativa se
presenta en diferentes campos de investigaci\'on como son astrof\'{\i}sica relativista, cosmolog\'{\i}a, geof\'{\i}sica,
entre otros. Precisamente debido a su gran aplicabilidad en diferentes y variados campos de investigaci\'on, actualmente la
transferencia radiativa constituye en s\'{\i} misma un campo de investigaci\'on a nivel an\'alitico y computacional; surge
también un creciente interés en el estudio de los aspectos relativistas del transporte de radiaci\'on debido a la necesidad
de considerarlos en el modelamiento y estudio de ambientes realistas, dada su gran influencia en tales modelos, como lo son
el estudio de atm\'osferas terrestres y/o estelares, o la corroboraci\'on de teor\'{\i}as acerca de la radiaci\'on detectada
por los telescopios de rayos-X CHANDRA y XMM-NEWTON. Aunque existe excelente bibliograf\'{\i}a sobre transferencia radiativa
\cite{mihalas1,chandrase1,rybicki1,pomraning}, el car\'acter covariante de la misma es apenas mencionado, y los c\'alculos
intermedios para su respectiva determinación o demostraci\'on no son desarrollados expl\'{\i}citamente, raz\'on por la
cu\'al el presente trabajo resulta ser un complemento a dicha bibliograf\'{\i}a, al desarrollar expl\'{\i}citamente los
c\'alculos, adem\'as de discutir acerca de las restricciones que se presentan en el transporte relativista de
radiaci\'on y como a pesar de la covarianza, resultan notorios los efectos relativistas especiales.\\

El presente trabajo se desarrolla comenzando con una introducci\'on a la transferencia radiativa (secci\'on \ref{clasica}),
donde se explica la metodología de obtención de la ecuaci\'on cl\'asica y las limitaciones que son impuestas en el estudio
del transporte de radiaci\'on no polarizada; posteriormente en la secci\'on \ref{especial} se discute la generalizaci\'on de
la ecuaci\'on covariante de Boltzmann y los efectos relativistas especiales que se presentan al tener en cuenta el
movimiento de la materia con la cual interactúa la radiaci\'on. Por último, en la secci\'on \ref{appen2} se realizan los
c\'alculos intermedios para mostrar el carácter invariante relativista especial de la ecuaci\'on de transferencia radiativa
y se distinguen los efectos relativistas que se presentan.
\section{Transferencia Radiativa Cl\'asica}\label{clasica}
La transferencia radiativa (\'o transporte de radiaci\'on) describe la interacci\'on de la radiaci\'on al propagarse por un
medio, teniendo en cuenta el punto de vista cu\'antico, en t\'erminos de cantidades macrosc\'opicas. Por lo tanto, la
transferencia radiativa proporciona informaci\'on de los efectos macrosc\'opicos de procesos cu\'anticos entre la
radiaci\'on y la materia \cite{ribicky,mihalas1,rtcnuestro}.\\

\begin{figure}[!h]
\begin{center}
\includegraphics[scale=0.5]{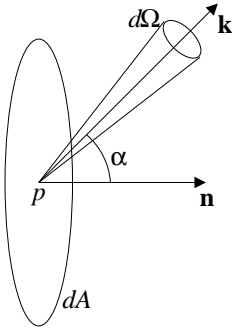}
\end{center}
\caption{Transporte de radiaci\'on en el punto $p$ del elemento de \'area $%
dA $}
\label{base11}
\end{figure}

Para el estudio del proceso de transporte de radiaci\'on se determina la cantidad de energ\'{\i}a $dE_{\nu}$ por unidad de
frecuencia $d\nu$, que est\'a atravesando un elemento de superficie $dA$ con vector normal $\boldsymbol{\mathbf{n}}$, en el
punto $p$ (tal como se muestra en la Figura \ref{base11}), en un intervalo de tiempo $dt$, donde la radiaci\'on est\'a
contenida en un \'angulo s\'olido $d\Omega$ alrededor de la direcci\'on de propagaci\'on $\boldsymbol{\mathbf{k}}$
\cite{aproxirayo}, es decir,

\begin{eqnarray}
dE_{\nu}&=I_{\nu}\hspace{0.15em}dA\hspace{0.15em}dt\hspace{0.15em}d\Omega%
\hspace{0.15em}d\nu\hspace{0.15em}\boldsymbol{\mathbf{n}}\centerdot%
\boldsymbol{\mathbf{k}}, \\
&=I_{\nu}\hspace{0.15em}dA\hspace{0.15em}dt\hspace{0.15em}d\Omega\hspace{%
0.15em}d\nu\hspace{0.15em}\cos(\alpha),  \notag
\end{eqnarray}

donde la \emph{intensidad espec\'{\i}fica} $I_{\nu}$ es una variable que da informaci\'on de los efectos macrosc\'opicos de
la interacci\'on de la
radiaci\'on (i.e., la energ\'{\i}a), al fluir a trav\'es de la materia \cite%
{mihalas1,chandrase1,rybicki1}.
Debido a que esta variaci\'on $dE_{\nu}$ depende del punto, direcci\'on,
frecuencia y tiempo de interacci\'on, entonces la intensidad espec\'{\i}fica
se podr\'a escribir como
\begin{equation}
I_{\nu}\equiv I(\boldsymbol{\mathbf{r}},\boldsymbol{\mathbf{k}};t,\nu),
\end{equation}
es decir, la intensidad espec\'{\i}fica va a dependerá de la posici\'on ($\boldsymbol{\mathbf{r}}$), del momento del fot\'on
[$\boldsymbol{\mathbf{p}}=(h\nu/c)\boldsymbol{\mathbf{k}}$] y del tiempo de interacci\'on ($t$). Así la transferencia
radiativa se puede ver como el cambio en la funci\'on de distribuci\'on
$f(\boldsymbol{\mathbf{r}},\boldsymbol{\mathbf{p}};t)$ de fotones al recorrer el espacio de fase en un tiempo ($t$) \cite{pieref}.\\

La manera de relacionar la intensidad espec\'{\i}fica con la funci\'on de distribuci\'on viene dada por
\begin{equation}  \label{ifunc}
I(\boldsymbol{\mathbf{r}},\boldsymbol{\mathbf{k}};t,\nu)=\frac{2 h {\nu}^3}{%
c^2}f(\boldsymbol{\mathbf{r}},\boldsymbol{\mathbf{p}};t),
\end{equation}
donde la multiplicación por un factor de $2$ corresponde a los dos posibles estados independientes de polarizaci\'on del
campo de radiaci\'on \cite{padmana1}. Por lo tanto, la variaci\'on de la intensidad espec\'{\i}fica $I_{\nu}$ \'o
equivalentemente el cambio de la funci\'on de distribuci\'on $f(\boldsymbol{\mathbf{r}}, \boldsymbol{\mathbf{p}};t)$
representa la manifestaci\'on
macrosc\'opica del transporte de radiaci\'on a trav\'es de la materia.\\

\subsection{Ecuaci\'on de Transferencia Radiativa}

\begin{figure}[!h]
\begin{center}
\includegraphics[scale=0.45]{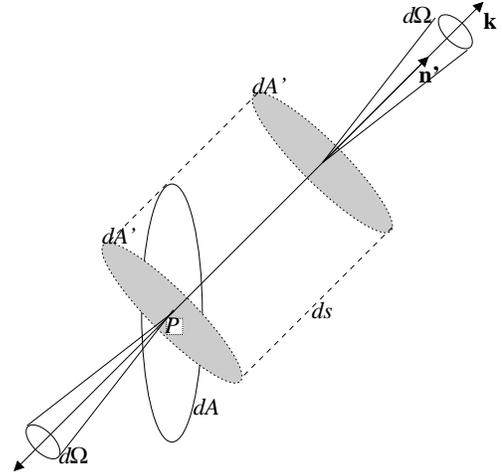}
\end{center}
\caption{Trayectoria de rayo de radiaci\'on}
\label{cilindro}
\end{figure}

Para determinar la variaci\'on de la intensidad espec\'{\i}fica al propagarse por un medio, se analiza el cambio de
energ\'{\i}a en un rayo debido a interacciones al recorrer una distancia $ds$. Por esta raz\'on, es necesario considerar un
cilindro de \'area $dA^{\prime}$ (en las ``tapas'') y longitud infinitesimal $ds$, alrededor del rayo (Figura
\ref{cilindro}) \cite{pieref2}, y se calcula el cambio del flujo de energ\'{\i}a del campo de radiaci\'on entre el haz
incidente al cilindro y el haz emergente de este \cite[P\'ags. 333 - 335]{mihalas1}.

Debido a que en general la radiaci\'on no se propaga en la misma direcci\'on a la normal al elemento de \'area $dA$ de la
Figura \ref{base11}, se considera un cilindro en direcci\'on $\boldsymbol{\mathbf{n^{\prime}}}$, con
\'area $dA^{\prime}$ (en las tapas del cilindro) y longitud $ds$, donde $%
dA^{\prime}=dA\hspace{0.12em}\cos(\alpha)$ es la proyecci\'on del elemento
de \'area del material sobre el plano definido por el vector normal $%
\boldsymbol{\mathbf{n^{\prime}}}$, de manera que los elementos de \'area $dA$
y $dA^{\prime}$ est\'an relacionados seg\'un la direcci\'on de propagaci\'on
de la radiaci\'on.\newline

De esta forma la variaci\'on de la intensidad espec\'{\i}fica al recorrer una distancia $ds$, debido a las interacciones
$g(\boldsymbol{\mathbf{r}},\boldsymbol{\mathbf{k}};t,\nu)$ está dada por
\begin{widetext}
\begin{equation}\label{eqrt}
\frac{d I(\boldsymbol{\mathbf{r}},\boldsymbol{\mathbf{k}};t,\nu)}{ds}\equiv\frac{d \boldsymbol{\mathbf{r}}}{ds}\cdot\frac{\partial I(\boldsymbol{\mathbf{r}},\boldsymbol{\mathbf{k}};t,\nu)}{\partial \boldsymbol{\mathbf{r}}}+\frac{d \boldsymbol{\mathbf{k}}}{ds}\cdot\frac{\partial I(\boldsymbol{\mathbf{r}},\boldsymbol{\mathbf{k}};t,\nu)}{\partial \boldsymbol{\mathbf{k}}}+\frac{dt}{ds}\frac{\partial I(\boldsymbol{\mathbf{r}},\boldsymbol{\mathbf{k}};t,\nu)}{\partial t}+\frac{d \nu}{ds}\frac{\partial I(\boldsymbol{\mathbf{r}},\boldsymbol{\mathbf{k}};t,\nu)}{\partial \nu}=g(\boldsymbol{\mathbf{r}},\boldsymbol{\mathbf{k}};t,\nu),
\end{equation}
\end{widetext}
donde el significado f\'{\i}sico de cada uno de los t\'erminos de la ecuaci\'on \eqref{eqrt} es:
($d\boldsymbol{\mathbf{r}}/ds$) es la direcci\'on de propagaci\'on del campo de radiaci\'on (i.e.,
$\boldsymbol{\mathbf{k}}$); ($d\boldsymbol{\mathbf{k}}/ds$) es el cambio en la direcci\'on de propagaci\'on en el momento en
que es transportada; ($dt/ds$) expresa el inverso de la velocidad de los fotones \cite{velocidad}; y ($d\nu/ds$) representa
el corrimiento en la frecuencia del campo a medida que se desplaza una distancia $ds$. El t\'ermino
$g(\boldsymbol{\mathbf{r}},\boldsymbol{\mathbf{k}};t,\nu)$ representa las interacciones entre la radiaci\'on y la materia, y
por lo tanto dichas interacciones son simplemente un tipo o combinaciones de
procesos como absorci\'on, emisi\'on y/o dispersi\'on (ver secci\'on \ref%
{inter} para los diferentes tipos de interacci\'on). De la relaci\'on %
\eqref{eqrt} se puede ver que al no presentarse interacci\'on (i.e., $g(%
\boldsymbol{\mathbf{r}},\boldsymbol{\mathbf{k}};t,\nu)=0$) la intensidad
espec\'{\i}fica es constante a lo largo de la trayectoria del rayo.\newline

La validez de la ecuaci\'on de transferencia \eqref{eqrt}, se restringe a
tratar al campo de radiaci\'on como una colecci\'on de part\'{\i}culas
cl\'asicas puntuales, donde el car\'acter cu\'antico de la descripci\'on se
presenta en la interacci\'on entre la radiaci\'on y la materia (e.g.,
absorci\'on de fotones por \'atomos, que resulta en \'atomos excitados o
ionizados) y no en cuantizar el campo electromagn\'etico \cite{qed}. Por
otro lado, al asumir que el transporte de radiaci\'on se realiza por medio
de rayos, implica que \'esta descripci\'on olvida efectos tales como
difracci\'on, interferencia y reflexi\'on ya que estos son efectos
ondulatorios de la radiaci\'on y no se pueden describir en esta
aproximaci\'on \cite[P\'ags. 47 - 49]{pomraning}.\newline

Debido a las limitaciones al despreciar efectos refractivos, el t\'ermino
referente al cambio en la direcci\'on de propagaci\'on se omite, aunque este
debe ser incluido cuando se utilicen coordenadas curvil\'{\i}neas \'o se
presente doblamiento gravitacional de la luz, as\'{\i} el transporte de
radiaci\'on sea en un medio no refractivo \cite{limi}. Adem\'as de la
anterior limitaci\'on, en transferencia radiativa cl\'asica se asume que la
materia con la cual esta interactuando el campo de radiaci\'on se encuentra
en reposo y no hay presentes campos gravitacionales intensos, es decir, no
se tienen en cuenta efectos relativistas especiales ni generales, de manera
que no hay corrimiento en la frecuencia a medida que el fot\'on se desplaza.%
\newline

Por lo tanto la ecuaci\'on de transferencia radiativa $d I_{\nu}/ds$
[ecuaci\'on \eqref{eqrt}] para un medio no refractivo, sin efectos
relativistas y en coordenadas curvil\'{\i}neas es
\begin{equation}  \label{dI/ds}
\frac{d I_{\nu}}{ds}=\frac{1}{c}\frac{\partial I_{\nu}}{\partial t}+\biggl\{%
\bigl\{\boldsymbol{\mathbf{k}}\centerdot\boldsymbol{\nabla}_{\boldsymbol{%
\mathbf{r}}}\bigr\}+\bigl\{\frac{d\text{\textbf{k}}}{ds}\centerdot%
\boldsymbol{\nabla}_{\boldsymbol{\mathbf{k}}}\bigr\} \biggr\}I_{\nu}=g_{\nu}(%
\boldsymbol{\mathbf{r}},\boldsymbol{\mathbf{k}};t),
\end{equation}
donde los t\'erminos $\boldsymbol{\nabla}_{\boldsymbol{\mathbf{r}}}$ y $%
\boldsymbol{\nabla}_{\boldsymbol{\mathbf{k}}}$ denotan diferenciaci\'on con
respecto a la posici\'on y a los cosenos directores de $\boldsymbol{\mathbf{k%
}}$, respectivamente.\newline

Como las coordenadas rectangulares son un caso especial de las coordenadas
curvil\'{\i}neas, la expresi\'on \eqref{dI/ds} en ese sistema de coordenadas
omite el tercer t\'ermino, ya que la direcci\'on de la base no cambia a lo
largo de la trayectoria dada por $ds$, por lo tanto la conocida ecuaci\'on
de transferencia radiativa cl\'asica en coordenadas rectangulares viene dada
por
\begin{equation}  \label{dI/dsrec}
\frac{d I_{\nu}}{ds}\hspace{0.15em}=\hspace{0.15em}\frac{1}{c}\hspace{0.15em}%
\frac{\partial I_{\nu}}{\partial t}\hspace{0.15em}+\hspace{0.15em}\bigl\{%
\boldsymbol{\mathbf{k}}\hspace{0.15em}\centerdot\hspace{0.15em}\boldsymbol{%
\nabla}_{\boldsymbol{\mathbf{r}}}\bigr\} I_{\nu}=g_{\nu}(\boldsymbol{\mathbf{%
r}},\boldsymbol{\mathbf{k}};t).
\end{equation}

\subsection{Interacci\'on}

\label{inter} Debido a que la transferencia radiativa da los aspectos
macrosc\'opicos de la interacci\'on cu\'antica entre la radiaci\'on con la
materia, \'esta es insuficiente para describir interacciones entre los
fotones, y las interacciones que se presentan se restringen al caso que la
trayectoria seguida por el fot\'on entre interacciones es una linea recta y
se desplaza con una velocidad igual a la del vac\'{\i}o $c$, i.e., se asume
a la materia como un \emph{polvo de electrones y \'atomos}, y por lo tanto
las interacciones son simplemente un tipo o combinaciones de procesos como:

\begin{itemize}
\item Emisi\'on Estimulada.

\item Emisi\'on Espont\'anea.

\item Absorci\'on.

\item Dispersi\'on.
\end{itemize}

donde la diferencia entre la emisi\'on espont\'anea y la estimulada, radica
en que la primera es independiente del campo de radiaci\'on, por lo tanto es
simplemente una fuente de energ\'{\i}a. Cuando los fotones no siguen
trayectorias rectas entre las colisiones, adem\'as de tener en cuenta las
anteriores interacciones se debe incluir efectos de refracci\'on, donde el
\'{\i}ndice de refracci\'on var\'{\i}e continuamente con respecto a la
posici\'on, y la trayectoria del fot\'on es una curva \cite[Cap. 5]%
{pomraning}.\newline

Aunque la transferencia radiativa sea una aproximaci\'on al estudio de la
interacci\'on de la radiaci\'on con la materia, esta es una gran herramienta
que permite estudiar ambientes y condiciones que no se pueden realizar en un
laboratorio terrestre. Algunas de sus aplicaciones en astrof\'{\i}sica son,
el estudio y modelamiento de la transferencia radiativa a trav\'es de
atm\'osferas estelares, de galaxias llenas de polvo (ver \cite{aplica} y
referencias contenidas ah\'{\i}), y en un proceso de acreci\'on \cite%
{shapiro1973,ti1998}; y en el campo cosmol\'ogico e.g. son el estudio del
efecto Sunyaev - Zel\'{}dovich \cite{cosmo}.

\section{Transferencia Radiativa Relativista Especial}

\label{especial} Debido a que en general la materia con la cual interact\'ua
la radiaci\'on no esta en reposo, para describir completamente la
transferencia radiativa en estos casos, se debe incluir los efectos que
induce dicho movimiento. Al tener en cuenta el que la materia est\'e en
movimiento constante, se debe hacer referencia con respecto a cual
observador se esta analizando la transferencia de radiaci\'on. Por ello se
considera dos observadores, uno Euleriano (laboratorio) y uno Lagrangiano
(com\'ovil, i.e., movi\'endose atado a la materia), de manera equivalente a
como se definen en mec\'anica de fluidos \cite{landau3}. Ya que estos dos
observadores se relacionan v\'{\i}a las transformaciones de Lorentz y
partiendo de la secci\'on \ref{clasica}, en la cual se analizo la
transferencia radiativa cuando la materia estaba en reposo con respecto a un
observador, i.e., la determinada por un marco (u observador) Lagrangiano, se
analiza la transferencia de radiaci\'on para un observador Euleriano.\newline

Como la f\'{\i}sica no puede depender del observador (i.e., el sistema de
referencia ver \cite[Cap. 4]{rybicki1} y \cite[P\'ags. 1 - 4]{schutz2}), se
debe escribir la ecuaci\'on de transferencia radiativa de manera covariante,
es decir, que la forma funcional sea invariante bajo transformaciones de
Lorentz. Para dicho prop\'osito y recurriendo a que la transferencia
radiativa es un caso particular de la f\'{\i}sica cin\'etica, se utiliza la
generalizaci\'on covariante de la ecuaci\'on de Boltzmann \cite[P\'ags. 418
- 419]{mihalas1}
\begin{equation}  \label{covarboltz}
\frac{df(\boldsymbol{\mathbb{X}},\boldsymbol{\mathbb{P}})}{d\tau}\equiv\frac{%
d x^{\mu}}{d\tau}\frac{\partial f(\boldsymbol{\mathbb{X}},\boldsymbol{%
\mathbb{P}})}{\partial x^{\mu}}+\frac{d p^{\mu}}{d \tau}\frac{\partial f(%
\boldsymbol{\mathbb{X}},\boldsymbol{\mathbb{P}})}{\partial p^{\mu}}=G(%
\boldsymbol{\mathbb{X}},\boldsymbol{\mathbb{P}}),
\end{equation}
donde $x^{\mu}$ y $p^{\mu}$ son las componentes de los correspondientes
cuadri-posici\'on $\boldsymbol{\mathbb{X}}$ (c-posici\'on) y cuadri-momento $%
\boldsymbol{\mathbb{P}}$ (c-momento) de la part\'{\i}cula (fot\'on); y $\tau$
es el par\'ametro af\'{\i}n que usualmente es el tiempo propio.\newline

La validez de la ecuaci\'on \eqref{covarboltz} impl\'{\i}citamente asume que
la funci\'on de distribuci\'on $f(\boldsymbol{\mathbb{X}},\boldsymbol{%
\mathbb{P}})$ es un invariante relativista. Al expandir expl\'{\i}citamente
la anterior ecuaci\'on, hay una dependencia de ocho par\'ametros (las cuatro
componentes de la c-posici\'on m\'as las cuatro componentes del c-momento)
que establece que la ecuaci\'on covariante de Boltzmann se desarrolla en un
espacio ocho dimensional, pero debido a que la norma del c-momento siempre
es una caracter\'{\i}stica inherente del sistema (proporcional a la masa en
reposo $m_{0}$, la cual es un invariante relativista) establece una ligadura
$\boldsymbol{\mathbb{P}}^2\equiv p^{\mu}p_{\mu}=-m_{0}^2c^2$ \cite[P\'ags.
45 - 47]{schutz2}, de manera que solo tres de las cuatro componentes de
c-momento son independientes, por tal raz\'on la ecuaci\'on %
\eqref{covarboltz} al igual que su versi\'on cl\'asica \eqref{eqrt} se
restringen a un espacio siete dimensional.\newline

Como las part\'{\i}culas que componen el campo de radiaci\'on son fotones
(part\'{\i}culas con masa en reposo cero) cuya velocidad es $c$ y tiempo
propio $\tau$ nulo, en la ecuaci\'on \eqref{covarboltz} el par\'ametro af%
\'{\i}n tiempo propio ya no es una variable \'util y por ello se emplea un
nuevo par\'ametro af\'{\i}n $\eta$, el cual es la longitud de trayectoria
recorrida del fot\'on. Por lo tanto, la visi\'on cin\'etica covariante de la
ecuaci\'on de transferencia radiativa es
\begin{equation}  \label{covarrt}
\frac{d x^{\mu}}{d\eta}\hspace{0.15em}\frac{\partial f(\boldsymbol{\mathbb{X}%
},\boldsymbol{\mathbb{P}})}{\partial x^{\mu}}+\frac{d p^{\mu}}{d \eta}\frac{%
\partial f(\boldsymbol{\mathbb{X}},\boldsymbol{\mathbb{P}})}{\partial p^{\mu}%
}\hspace{0.15em} = \hspace{0.15em}G(\boldsymbol{\mathbb{X}},\boldsymbol{%
\mathbb{P}}),
\end{equation}
donde el c-momento $\boldsymbol{\mathbb{P}}$ del fot\'on se define como $%
p^{\mu}\equiv dx^{\mu}/d\eta$, con $\eta$ el par\'ametro af\'{\i}n longitud
de trayectoria recorrida del fot\'on.\newline

Debido a que la relatividad especial \'unicamente ingresa los efectos
cinem\'aticos entre observadores que se est\'en moviendo a velocidad
constante uno con respecto al otro, las condiciones impuestas en la
secci\'on de transferencia radiativa cl\'asica \ref{clasica} acerca de que
no hay fuerzas que afecten la trayectoria del fot\'on (conservaci\'on del
momento), tambi\'en son validas en la visi\'on relativista especial y sus
efectos ser\'an que el fot\'on se propague en una linea recta, pero con
direcci\'on, longitud de onda y frecuencia (por la constancia de la
velocidad de la luz, el cambio en la frecuencia se debe compensar con el
cambio en la longitud de onda) que dependen del observador, por lo tanto el
t\'ermino ${d p^{\mu}}/{d \eta}$ en \eqref{covarrt} es cero (i.e., la
conservaci\'on del c-momento). Entonces la ecuaci\'on covariante de la
transferencia radiativa no polarizada en un medio no refractivo es
\begin{equation}  \label{covarrtrs}
p^{\mu}\hspace{0.15em}\frac{\partial f(\boldsymbol{\mathbb{X}},\boldsymbol{%
\mathbb{P}})}{\partial x^{\mu}}\hspace{0.15em} = \hspace{0.15em}G(%
\boldsymbol{\mathbb{X}},\boldsymbol{\mathbb{P}}).
\end{equation}

Como se mencion\'o, el car\'acter covariante bajo \emph{boost} de Lorentz de
una ecuaci\'on significa que esta sea invariante en su forma para
observadores que est\'an relacionados por una transformaci\'on de Lorentz
(boost). Por lo tanto, al demostrar que la ecuaci\'on de transferencia
radiativa, desde el punto de vista cin\'etico es covariante [ecuaci\'on %
\eqref{covarrtrs}] y teniendo en cuenta las mismas condiciones tanto
cl\'asicas como relativistas especiales acerca de que no hay fuerzas que
afecten la trayectoria del fot\'on, i.e., la correspondiente conservaci\'on
del momento (desde el punto de vista cl\'asico) y el c-momento (desde el
punto de vista relativista), al expandir el \'{\i}ndice $\mu$, y como las
componentes del c-momento $\boldsymbol{\mathbb{P}}$ son $p^{\mu}=h\nu(1,%
\{1/c\}\boldsymbol{\mathbf{k}})$, se obtiene la ecuaci\'on de transferencia
radiativa cl\'asica \eqref{dI/dsrec}. Los efectos cinem\'aticos en este
momento no aparecen, sino que ellos van a hacerse relevantes al combinar los
dos observadores.

\subsection{Efectos Relativistas}

\label{efectosrela} Como se mencion\'o en la secci\'on \ref{especial} la
\'unica manera de distinguir los efectos relativistas, es comparar lo que
dos observadores ven cuando cada uno se encuentra en un estado movimiento
diferente. Para el caso relativista especial los efectos observados son:

\begin{itemize}
\item Aberraci\'on: La direcci\'on de propagaci\'on del fot\'on depende del
estado de movimiento del observador.

\item Corrimiento Doppler: Corrimiento en la frecuencia y por lo tanto en la
longitud de onda para compensar la constancia de la velocidad del fot\'on.

\item Advectivo: Arrastre del fot\'on por un medio en movimiento. En este
punto se debe aclarar que el arrastre del fot\'on no producir\'a que su
velocidad de propagaci\'on sea mayor a la velocidad de la luz en el vac\'{\i}%
o ($c$), ya que al propagarse un campo de radiaci\'on a trav\'es de un medio
diferente al vac\'{\i}o (donde dicho medio se caracteriza por un \'{\i}ndice
de refracci\'on $n$, tal que $n>1$) su velocidad de propagaci\'on esta dada
por $v=c/n$, es decir, que disminuye con respecto a $c$.
\end{itemize}

\subsection{T\'ermino de interacci\'on $G(\boldsymbol{\mathbb{X}},%
\boldsymbol{\mathbb{P}})$}

\label{invariaintera} Debido a que los fen\'omenos de interacci\'on $G(%
\boldsymbol{\mathbb{X}},\boldsymbol{\mathbb{P}})$ no pueden depender del
sistema de referencia, ya que implicar\'{\i}a que mientras un observador
``ve'' que se presenta una absorci\'on, otro observador que esta relacionado
al anterior por medio de una transformaci\'on de Lorentz observar\'{\i}a que
la interacci\'on con el campo de radiaci\'on es una emisi\'on estimulada.
Por lo tanto por el primer postulado de la relatividad \emph{La F\'{\i}sica
no depende del observador ni de su estado cin\'ematico} \cite{schutz2}, se
entiende que los procesos f\'{\i}sicos que ocurren en la naturaleza son
validos para cualquier observador inercial, y lo que sucede es que cada
observador inercial observa el mismo proceso f\'{\i}sico pero con diferente
valor num\'erico y direcci\'on, por consiguiente es cuando los efectos
cinem\'aticos descritos en la secci\'on anterior afectan los t\'erminos de
interacci\'on, en el sentido que los mismos procesos f\'{\i}sicos son
observados por distintos marcos inerciales pero que suceden a distintos
valores de energ\'{\i}a y de direcci\'on de propagaci\'on para cada
observador inercial.

\section{Invariancia de la ecuaci\'on de transferencia radiativa}

\label{appen2} La ecuaci\'on de transferencia radiativa cl\'asica es
invariante bajo transformaciones de Lorentz, por lo tanto para mostrar dicha
invariancia se tiene que la ecuaci\'on de transferencia del marco com\'ovil ($%
\Sigma^{\prime}$)
\begin{equation}\label{eqreferi}
\frac{1}{c}\frac{\partial I^{\prime}}{\partial t^{\prime}}\hspace{0.15em}+%
\hspace{0.15em}\boldsymbol{{\mathbf{k}^{\prime}}}\hspace{0.15em}\centerdot%
\hspace{0.15em}\boldsymbol{\nabla^{\prime}} I^{\prime}\hspace{0.15em}=%
\hspace{0.15em} g^{\prime},
\end{equation}
debe ser funcionalmente igual al del marco de laboratorio ($\Sigma$)
\begin{equation}
\frac{1}{c}\frac{\partial I}{\partial t}\hspace{0.15em}+\hspace{0.15em}%
\boldsymbol{{\mathbf{k}}}\hspace{0.15em}\centerdot\hspace{0.15em}\boldsymbol{%
\nabla} I \hspace{0.15em}=\hspace{0.15em} g,
\end{equation}
cuando los dos marcos ($\Sigma^{\prime}$ y $\Sigma$) est\'an relacionados
por un boost en direcci\'on arbitraria. La transformaci\'on de Lorentz que
relaciona dos marcos inerciales con velocidad relativa ($\boldsymbol{{%
\mathbf{v}}}$) entre s\'{\i}, es la transformaci\'on de Lorentz (boost) de
velocidad $\boldsymbol{{\mathbf{\beta}}}$ (donde $\boldsymbol{{\mathbf{\beta}%
}}\equiv\frac{\boldsymbol{{\mathbf{v}}}}{c}$ y $\beta\equiv\mid\boldsymbol{%
\mathbf{\beta}}\mid$) y direcci\'on arbitraria \cite{notacionmihalas}
\begin{equation}  \label{boost}
\Lambda^{\mu^{\prime}}_{\nu}=
\begin{pmatrix}
\gamma & \hspace{0.15em} & \hspace{0.15em}-\gamma\hspace{0.15em}\boldsymbol{{%
\mathbf{\beta}}} \\
-\gamma\hspace{0.15em}\boldsymbol{{\mathbf{\beta}}} & \hspace{0.15em} &
\hspace{0.15em}\boldsymbol{\mathbb{I}}\hspace{0.15em}+\hspace{0.15em}(\gamma
- 1)\hspace{0.15em}\beta^{-2}\hspace{0.15em}\boldsymbol{{\mathbf{\beta}}}%
\hspace{0.15em}\boldsymbol{{\mathbf{\beta}}}%
\end{pmatrix}%
,
\end{equation}
por completez de la demostraci\'on se escribe expl\'{\i}citamente $%
\partial_{\mu}$ y $\mathrm{P}^{\mu}$, los cuales son:
\begin{equation}  \label{divergencia}
\partial_{\mu}=\frac{\partial}{\partial x^{\mu}}=\left(\frac{1}{c}\frac{%
\partial}{\partial t},\boldsymbol{\nabla}\right),
\end{equation}
\begin{equation}  \label{c-vmomentofoton}
\mathrm{P}^{\mu}=\left(\frac{E}{c},\boldsymbol{{\mathbf{P}}}\right)=\frac{%
h\nu}{c}\left(1,\boldsymbol{{\mathbf{k}}}\right).
\end{equation}

Recordando que un c-vector contravariante $x^{\mu^{\prime}}$ en el sistema
de referencia $\Sigma^{\prime}$ est\'a relacionado con su correspondiente
c-vector covariante $x^\nu$ en el sistema de referencia $\Sigma$ por
\begin{equation}
x^{\mu^{\prime}}=\Lambda^{\mu^{\prime}} _\nu x^\nu,
\end{equation}
y para c-vectores covariantes
\begin{equation}
x_{\mu^{\prime}}=\Lambda_{\mu^{\prime}} ^\nu x_\nu,
\end{equation}
donde $\Lambda^{\mu^{\prime}}_\nu
\Lambda_{\mu^{\prime}}^\alpha=\delta^\alpha_\nu $ es el s\'{\i}mbolo delta
de Kronecker.\newline

Retomando la transformaci\'on de Lorentz \eqref{boost}, y reemplazando los
t\'erminos $\boldsymbol{{\mathbf{\beta}}}$ por una notaci\'on vectorial
m\'as conocida $\boldsymbol{{\mathbf{\beta}}}$, las transformaciones entre
los marcos $\Sigma - \Sigma^{\prime}$ y $\Sigma^{\prime} - \Sigma$,
respectivamente son
\begin{equation}
\Lambda^{\mu^{\prime}}_\nu =
\begin{pmatrix}
\gamma & -\gamma\boldsymbol{{\mathbf{\beta}}} \\
-\gamma\boldsymbol{{\mathbf{\beta}}} & I+(\gamma-1)\beta^{-2}\boldsymbol{{%
\mathbf{\beta}}}\boldsymbol{{\mathbf{\beta}}}%
\end{pmatrix}%
\end{equation}
y
\begin{equation}
\Lambda_{\mu^{\prime}}^\nu=
\begin{pmatrix}
\gamma & \gamma\boldsymbol{{\mathbf{\beta}}} \\
\gamma\boldsymbol{{\mathbf{\beta}}} & I+(\gamma-1)\beta^{-2}\boldsymbol{{%
\mathbf{\beta}}}\boldsymbol{{\mathbf{\beta}}}%
\end{pmatrix}%
,
\end{equation}
siendo $\boldsymbol{{\mathbf{\beta}}}\boldsymbol{{\mathbf{\beta}}}$ una
diada, $\gamma^2=(1-\beta^2)^{-1}$ y $\boldsymbol{{\mathbf{\beta}}}=\frac{%
\boldsymbol{{\mathbf{v}}}}{c}$.\newline

Para demostrar la invariancia de la ecuaci\'on de transferencia radiativa %
\eqref{dI/dsrec}, primero se debe transformar el c-vector divergencia
[ecuaci\'on \eqref{divergencia}] por medio de la transformaci\'on de
Lorentz, lo cual viene dado por
\begin{equation}
\partial_{\mu\prime}=\Lambda_{\mu\prime}^\alpha\partial_\alpha,
\end{equation}
obteniendo para cada una de sus componentes
\begin{equation}  \label{aaaa1}
\frac{1}{c}\frac{\partial}{\partial t^{\prime}}=\gamma\left(\frac{1}{c}\frac{%
\partial}{\partial t}+\boldsymbol{{\mathbf{\beta}}}\centerdot\boldsymbol{{%
\mathbf{\nabla}}}\right),
\end{equation}
\begin{equation}  \label{aa1}
\boldsymbol{{\mathbf{\nabla^{\prime}}}}=\gamma\boldsymbol{{\mathbf{\beta}}}%
\frac{1}{c}\frac{\partial}{\partial t}+\boldsymbol{{\mathbf{\nabla}}}%
+(\gamma-1)\beta^{-2}\boldsymbol{{\mathbf{\beta}}}\boldsymbol{{\mathbf{\beta}%
}}\centerdot\boldsymbol{{\mathbf{\nabla}}},
\end{equation}
y segundo se debe realizar la transformaci\'on de Lorentz del c-vector
momento del fot\'on [ecuaci\'on \eqref{c-vmomentofoton}], obteniendo para
cada una de sus componentes
\begin{equation}  \label{aaaa2}
\nu^{\prime}=\gamma\nu\left(1-\boldsymbol{{\mathbf{\beta}}}\centerdot%
\boldsymbol{{\mathbf{k}}}\right),
\end{equation}
\begin{equation}  \label{aaaa3}
\boldsymbol{{\mathbf{k^{\prime}}}}=\frac{\nu}{\nu^{\prime}}\left(-\gamma%
\boldsymbol{{\mathbf{\beta}}}+\boldsymbol{{\mathbf{k}}}+(\gamma-1)\beta^{-2}%
\boldsymbol{{\mathbf{\beta}}}\boldsymbol{{\mathbf{\beta}}}\centerdot%
\boldsymbol{{\mathbf{k}}}\right).
\end{equation}

Las expresiones desde la ecuaci\'on \eqref{aaaa1} hasta la ecuaci\'on %
\eqref{aaaa3}, son las necesarias para demostrar la invariancia de la
ecuaci\'on de transferencia radiativa, adicionalmente los efectos
relativistas corrimiento Doppler y aberraci\'on (secci\'on \ref{efectosrela}%
) se pueden observar en las transformaciones de cada una de las componentes
de c-vector momento del fot\'on, por lo tanto reagrupando y ordenando los
t\'erminos se obtiene que el factor de corrimiento entre las frecuencias $%
\nu^{\prime}$ y $\nu$ esta dado por
\begin{equation}  \label{a1}
\frac{\nu^{\prime}}{\nu}=\gamma(1-\boldsymbol{{\mathbf{\beta}}}\centerdot%
\boldsymbol{{\mathbf{k}}}),
\end{equation}
y utilizando $\beta^{-2}=\frac{\gamma^2}{\gamma^2-1}$ en la ecuaci\'on %
\eqref{aaaa3}, el efecto de aberraci\'on entre las direcciones de
propagaci\'on $\boldsymbol{{\mathbf{k^{\prime}}}}$ y $\boldsymbol{{\mathbf{k}%
}}$ viene dado por
\begin{equation}  \label{a2}
\boldsymbol{{\mathbf{k^{\prime}}}}=\frac{\nu}{\nu^{\prime}}\left\{%
\boldsymbol{{\mathbf{k}}}-\gamma\boldsymbol{{\mathbf{\beta}}}\left(1-\frac{%
\gamma\boldsymbol{{\mathbf{\beta}}}\centerdot\boldsymbol{{\mathbf{k}}}}{%
\gamma +1}\right)\right\},
\end{equation}
una visi\'on gr\'afica del comportamiento de estos efectos [ecuaciones %
\eqref{a1} y \eqref{a2}] se puede ver en el ap\'endice \ref{comportamiento}.%
\newline

Utilizando la relaci\'on de la intensidad invariante $\boldsymbol{{\mathbf{I}%
}}$ \cite[P\'ags. 413 - 414]{mihalas1} la cual viene dada por%
\begin{equation}  \label{a23}
\boldsymbol{{\mathbf{I}}}\equiv\frac{I^{\prime}}{\nu^{\prime 3}}=\frac{I}{%
\nu^3},
\end{equation}
y la relaci\'on entre las frecuencias $\nu$ y $\nu^{\prime}$ [ecuaci\'on %
\eqref{a1}], la expresi\'on entre las [ecuaci\'on \eqref{a23}] intensidades
en los marcos Euleriano y Lagrangiano, quedan determinadas como:
\begin{equation}  \label{a4}
I^{\prime}=\gamma^3(1-\boldsymbol{{\mathbf{\beta}}}\centerdot\boldsymbol{{%
\mathbf{k}}})^3I.
\end{equation}

Remplazando las cantidades en el marco com\'ovil por las correspondientes al
marco del laboratorio (las cuales se relacionan v\'{\i}a las
transformaciones de Lorentz), se tiene que
\begin{equation}  \label{a3}
\frac{1}{c}\frac{\partial}{\partial t^{\prime}}I^{\prime}+\boldsymbol{{%
\mathbf{k^{\prime}}}}\centerdot\boldsymbol{{\mathbf{\nabla}^{\prime}}}%
I^{\prime}=g^{\prime}
\end{equation}
Remplazando en la ecuaci\'on \eqref{a3} los resultados de \eqref{aa1}, %
\eqref{a2} y \eqref{a4} se tiene
\begin{widetext}
\begin{eqnarray}\label{a5}
&\gamma\left(\frac{1}{c}\frac{\partial}{\partial t}+\boldsymbol{{\mathbf{\beta}}}\centerdot\boldsymbol{{\mathbf{\nabla}}}\right)\left(\gamma^3(1-\boldsymbol{{\mathbf{\beta}}}\centerdot\boldsymbol{{\mathbf{k}}})^3\right)I+\frac{1}{\gamma(1-\boldsymbol{{\mathbf{\beta}}}\centerdot\boldsymbol{{\mathbf{k)}}}}\left(\boldsymbol{{\mathbf{k}}}-\gamma\boldsymbol{{\mathbf{\beta}}}\left(1-\frac{\gamma\boldsymbol{{\mathbf{\beta}}}\centerdot\boldsymbol{{\mathbf{k}}}}{\gamma+1}\right)\right)\centerdot\left(\boldsymbol{{\mathbf{\nabla}}}+\gamma\boldsymbol{{\mathbf{\beta}}}\left(\frac{1}{c}\frac{\partial}{\partial t}+\frac{\gamma}{\gamma+1}\boldsymbol{{\mathbf{\beta}}}\boldsymbol{{\mathbf{\beta}}}\centerdot\boldsymbol{{\mathbf{\nabla}}}\right)\right)\nonumber\\
&\left(\gamma^3(1-\boldsymbol{{\mathbf{\beta}}}\centerdot\boldsymbol{{\mathbf{k}}})^3\right)I=g^{\prime}.
\end{eqnarray}
\end{widetext}
Suponiendo que no se presentan efectos de refracci\'on o doblamiento
gravitacional de la luz (ya que en el marco com\'ovil se determina la
ecuaci\'on de transferencia radiativa sin doblamiento gravitacional de la
luz \'o refracci\'on, en el marco de laboratorio se debe observar lo mismo,
debido a que la f\'{\i}sica para ambos observadores debe ser la misma), las
variaciones de $\gamma(1-\boldsymbol{{\mathbf{\beta}}}\centerdot\boldsymbol{{%
\mathbf{k}}})$ son nulas, i. e., no se presentan. Por lo tanto factorizando
el t\'ermino $\gamma^2(1-\boldsymbol{{\mathbf{\beta}}}\centerdot\boldsymbol{{%
\mathbf{k}}})^2$ en la expresi\'on \eqref{a5} se encuentra
\begin{widetext}
\begin{equation}
\gamma^2(1-\boldsymbol{{\mathbf{\beta}}}\centerdot\boldsymbol{{\mathbf{k}}})\left(\frac{1}{c}\frac{\partial}{\partial t}+\boldsymbol{{\mathbf{\beta}}}\centerdot\boldsymbol{{\mathbf{\nabla}}}\right)I+\left(\boldsymbol{{\mathbf{k}}}-\gamma\boldsymbol{{\mathbf{\beta}}}\left(1-\frac{\gamma\boldsymbol{{\mathbf{\beta}}}\centerdot\boldsymbol{{\mathbf{k}}}}{\gamma+1}\right)\right)\centerdot\left(\boldsymbol{{\mathbf{\nabla}}}+\gamma\boldsymbol{{\mathbf{\beta}}}\left(\frac{1}{c}\frac{\partial}{\partial t}+\frac{\gamma}{\gamma+1}\boldsymbol{{\mathbf{\beta}}}\centerdot\boldsymbol{{\mathbf{\nabla}}}\right)\right)I=G,
\end{equation}
\end{widetext}
donde $G=\frac{g^{\prime}}{\gamma^2(1-\boldsymbol{{\mathbf{\beta}}}\centerdot%
\boldsymbol{{\mathbf{k}}})^2}$. Expandiendo los t\'erminos y agrup\'andolos
en un par\'entesis con un factor com\'un $\frac{1}{c}\frac{\partial I}{%
\partial t}$ y uno con $\boldsymbol{{\mathbf{\nabla}}}I$ se encuentra
\begin{widetext}
\begin{eqnarray}
&\frac{1}{c}\frac{\partial I}{\partial t}\left(\gamma^2-\gamma^2\boldsymbol{{\mathbf{\beta}}}\centerdot\boldsymbol{{\mathbf{k}}}+\gamma\boldsymbol{{\mathbf{k}}}\centerdot\boldsymbol{{\mathbf{\beta}}}-\gamma^2\beta^2+\frac{\gamma^3\boldsymbol{{\mathbf{\beta}}}\centerdot\boldsymbol{{\mathbf{k}}}\beta^2}{\gamma+1}\right)+\boldsymbol{{\mathbf{k}}}\centerdot\boldsymbol{{\mathbf{\nabla}}}I+\nonumber\\
&\left(\gamma^2-\gamma^2\boldsymbol{{\mathbf{\beta}}}\centerdot\boldsymbol{{\mathbf{k}}}+\frac{\gamma^2\boldsymbol{{\mathbf{k}}}\centerdot\boldsymbol{{\mathbf{\beta}}}}{\gamma+1}-\gamma+\frac{\gamma^2(\boldsymbol{{\mathbf{\beta}}}\centerdot\boldsymbol{{\mathbf{k}}})}{\gamma+1}-\frac{\gamma^3\beta^2}{\gamma+1}+\frac{\gamma^4\beta^2(\boldsymbol{{\mathbf{\beta}}}\centerdot\boldsymbol{{\mathbf{k}}})}{(\gamma+1)^2}\right)\boldsymbol{{\mathbf{\beta}}}\centerdot\boldsymbol{{\mathbf{\nabla}}}I=G.
\end{eqnarray}
\end{widetext}
Simplificando por separado el primero y el segundo par\'entesis se encuentra
\begin{equation}
\left(\gamma^2-\gamma^2\boldsymbol{{\mathbf{\beta}}}\centerdot\boldsymbol{{%
\mathbf{k}}}+\gamma\boldsymbol{{\mathbf{k}}}\centerdot\boldsymbol{{\mathbf{%
\beta}}}-\gamma^2\beta^2+\frac{\gamma^3\boldsymbol{{\mathbf{\beta}}}%
\centerdot\boldsymbol{{\mathbf{k}}}\beta^2}{\gamma+1}\right)=1,
\end{equation}
y
\begin{multline}
\gamma^2-\gamma^2\boldsymbol{{\mathbf{\beta}}}\centerdot\boldsymbol{{\mathbf{%
k}}}+\frac{\gamma^2\boldsymbol{{\mathbf{k}}}\centerdot\boldsymbol{{\mathbf{%
\beta}}}}{\gamma+1}-\gamma+\frac{\gamma^2(\boldsymbol{{\mathbf{\beta}}}%
\centerdot\boldsymbol{{\mathbf{k}}})}{\gamma+1} \\
-\frac{\gamma^3\beta^2}{\gamma+1}+\frac{\gamma^4\beta^2(\boldsymbol{{\mathbf{%
\beta}}}\centerdot\boldsymbol{{\mathbf{k}}})}{(\gamma+1)^2}=0,
\end{multline}
por lo tanto se obtiene la ecuaci\'on de transferencia radiativa en el marco
del laboratorio como
\begin{equation}
\frac{1}{c}\frac{\partial I}{\partial t}+\boldsymbol{{\mathbf{k}}}\centerdot%
\boldsymbol{{\mathbf{\nabla}}}I=G,
\end{equation}
la cual es funcionalmente id\'entica a la del marco com\'ovil \eqref{a3}. De
lo anterior se distingue que los efectos cinem\'aticos relativistas se
presentan en la transformaci\'on de la frecuencia y direcci\'on de
propagaci\'on, i.e., corrimiento Doppler y aberraci\'on.

La anterior demostraci\'on  \footnote{Otra manera de demostrar la 
invarianza de la ecuaci\'on de transferencia radiativa \eqref{eqreferi} 
aunque de manera m\'as directa y compacta es la consignada en el ap\'endice 
\ref{referi}.} muestra que tanto para un observador Euleriano como para 
un observador Lagrangiano (i.e., un marco de laboratorio y un marco com\'ovil, 
respectivamente) determinan el mismo tipo de ecuaci\'on de transferencia radiativa, 
sin nungun efecto aparente de su estado de movimiento. Por tal razon, en el estudio de procesos 
que involucran la transferencia radiativa en medios en movimiento, \'o observadores
movi\'endose se requiere de un marco mixto, es decir, un marco donde
las variables dependientes est\'en en el marco de laboratorio y las
variables independientes est\'en en el marco de com\'ovil.

\section{Conclusiones}
La demostraci\'on de la invariancia relativista especial de la ecuaci\'on de
transferencia radiativa, encuentra que aun cuando no existe ning\'un tipo de
interacci\'on entre la radiaci\'on y la materia, se presentan los efectos rela
tivistas de aberraci\'on y corrimiento Doppler, como consecuencia de que la ma
teria est\'e en movimiento y dichos efectos son independientes del tipo de inte
racci\'on entre la radiaci\'on y la materia; el presente trabajo sirve como com
plemento y aclaraci\'on del por qu\'e de la necesidad de marcos mixtos en el
tratamiento relativista de la ecuación de transferencia radiativa.

El efecto de corrimiento Doppler (factor de corrimiento) tiene un comportamiento irregular, ya que
a medida que la norma del boost tiende a su valor m\'aximo este va a cero cuando el ángulo
entre los observadores es cero, lo cual muestra que cuando un
observador va en la misma dirección de propagación de la
radiación y a medida que su velocidad aumenta, este sistema observa que la
frecuencia que mide y la que tiene la radiación tienden a ser las
mismas, esto se observa en la figura \ref{angles-ctes-1}.

El efecto de aberración tiene un valor límite $\pi$ a diferencia del factor de
corrimiento que puede variar hasta infinito. En la figura \ref{boost-ctes} se
observa que sin importar el ángulo entre los observadores a
medida que la norma del boost aumenta la aberración tiende a su valor
límite, siempre y cuando el ángulo entre los observadores sea
diferente de cero. Adicionalmente en la figura \ref{angles-ctes} se encuentra que
el efecto de aberraci\'on es independiente del boost cuando el \'angulo entre los
observadores es $0$ \'o $\pi$.
\renewcommand{\acknowledgmentsname}{AGRADECIMIENTOS}
\begin{acknowledgments}
Los autores desean agradecer al Departamento de F\'{\i}sica de la Universidad Nacional de Colombia  donde este trabajo fue desarrollado, adicionalmente desean agradecer por los valiosos comentarios por parte del refer\'{\i} que hicieron m\'as completo el presente trabajo.
\end{acknowledgments}

\renewcommand{\appendixname}{AP\'ENDICE} \appendix

\section{Comportamiento de los efectos de Corrimiento Doppler y Aberraci\'on}\label{comportamiento}
A continuaci\'on se presenta el comportamiento de los
efectos relativistas de corrimiento Doppler \eqref{a1} y aberraci\'on %
\eqref{a2}, con respecto al \'angulo entre la direcci\'on de propagaci\'on $%
\boldsymbol{\mathbf{k}}$ y el boost $\boldsymbol{\mathbf{\beta}}$.
\subsection{Corrimiento Doppler}
\begin{figure}[!h]
\begin{center}
\includegraphics[scale=1]{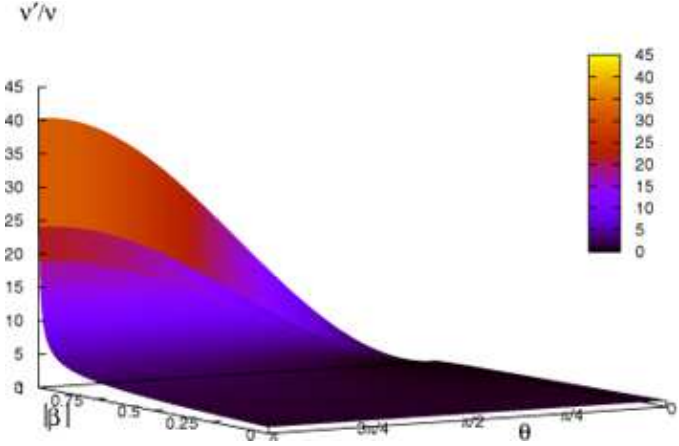}
\end{center}
\caption{Comportamiento del factor de corrimiento con respecto a la norma
del boost $\mid\boldsymbol{\mathbf{\protect\beta}}\mid$ y el \'angulo $%
\protect\theta$.}
\label{base-doppler}
\end{figure}

El factor de corrimiento [ecuaci\'on \eqref{a1}] est\'a dado por
\begin{equation}  \label{apendicea1}
\frac{\nu^{\prime}}{\nu}=\frac{1-\boldsymbol{{\mathbf{\beta}}}\centerdot%
\boldsymbol{{\mathbf{k}}}}{\sqrt{1-\beta^2}},
\end{equation}
donde el t\'ermino $\boldsymbol{{\mathbf{k}}}$ indica la direcci\'on de
propagaci\'on de la radiaci\'on. Debido a que en la ecuaci\'on %
\eqref{apendicea1} se presenta el producto escalar $\boldsymbol{{\mathbf{%
\beta}}}\centerdot\boldsymbol{{\mathbf{k}}}$, el factor de corrimiento
depende de las variables independientes $\beta\equiv\mid\boldsymbol{\mathbf{%
\beta}}\mid$ y $\theta$, siendo $\theta$ el \'angulo entre la direcci\'on de
propagaci\'on $\boldsymbol{\mathbf{k}}$ y el boost $\boldsymbol{\mathbf{\beta%
}}$, de manera que la expresi\'on \eqref{apendicea1} se puede escribir como
\begin{equation}  \label{apendicea2}
\frac{\nu^{\prime}}{\nu}=\frac{1-\beta \cos(\theta)}{\sqrt{1-\beta^2}},
\end{equation}
donde se ha tomado que la norma del vector direcci\'on de propagaci\'on $\mid%
\boldsymbol{\mathbf{k}}\mid$ es igual a la unidad \cite{confusion}. El
comportamiento del factor de corrimiento \eqref{apendicea2} se puede
observar en la figura \ref{base-doppler}.

\begin{figure}[!h]
\begin{center}
\includegraphics[scale=0.7]{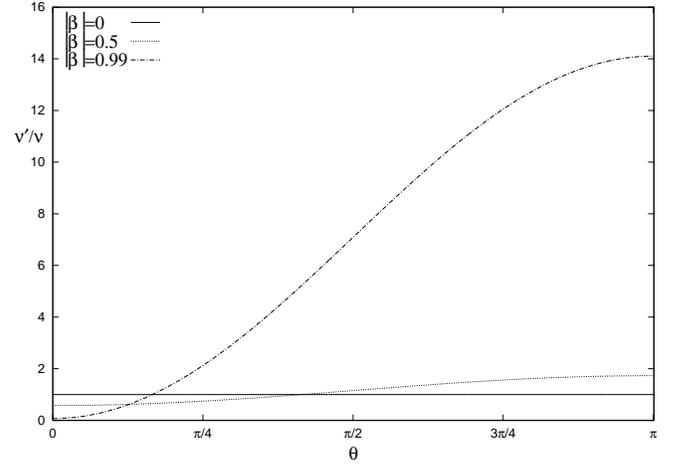}
\end{center}
\caption{Comportamiento del factor de corrimiento a boost constantes}
\label{boost-ctes-1}
\end{figure}

\begin{figure}[!h]
\begin{center}
\includegraphics[scale=0.7]{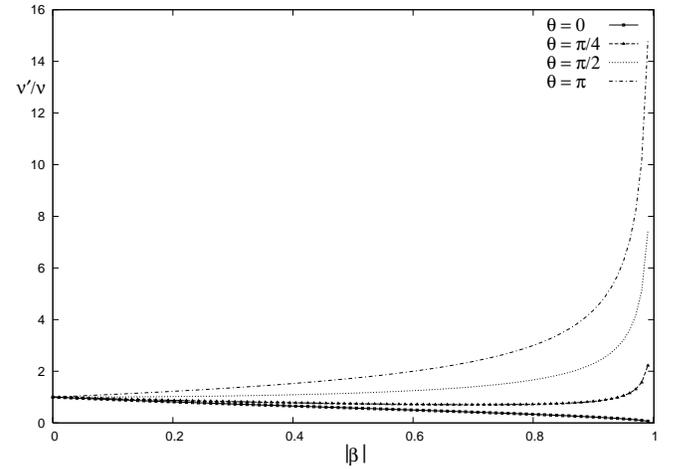}
\end{center}
\caption{Comportamiento del factor de corrimiento a \'angulos constantes}
\label{angles-ctes-1}
\end{figure}

Una visi\'on detallada del comportamiento del factor de corrimiento (figura %
\ref{base-doppler}) se puede observar en la figura \ref{boost-ctes-1} (boost
constantes en magnitud) y en la figura \ref{angles-ctes-1} (\'angulos
constantes).

En la figura \ref{boost-ctes-1} se observa como el factor de corrimiento para un boost igual a cero es igual a la unidad, ya
que esto significa que son dos sistemas que est\'an en reposo relativo entre si, pero que sus ejes pueden estar rotando un
\'angulo $\theta$ entre ellos. Para un boost diferente de cero (e.g., $\beta=0.5$ \'o $\beta=0.99$) se puede observar como
el factor de corrimiento no solo va a depender de la norma de la velocidad relativa entre los sistemas inerciales sino de su
direcci\'on de manera que el factor $\nu^{\prime}/\nu$ crece a medida que la norma del boost y su direcci\'on tienden hacia
sus valores extremos, respectivamente.

En la figura \ref{angles-ctes-1} se observa como el factor de corrimiento
para un \'angulo $\theta$ igual a cero tiende a cero, ya que esto significa
dos sistemas que se encuentran dirigidos en la misma direcci\'on, pero a
medida que aumenta la velocidad relativa entre ellos la frecuencia de la
radiaci\'on que ve uno de ellos tiende a ser nula, ya que uno de los
sistemas al aumentar su velocidad relativa respect\'o al otro sistema, se
aproximar\'{\i}a a la velocidad de propagaci\'on de la onda por lo que no ver\'{\i}a
ninguna onda, por lo tanto el factor de corrimiento entre los observadores
seria nulo. Para un \'angulo diferente de cero (e.g., $\theta=\pi/2$ \'o $%
\theta=\pi/4$) como el factor de corrimiento crece r\'apidamente a medida que
la velocidad relativa entre los sistemas tiende a la unidad.

\subsection{Aberraci\'on}

\begin{figure}[!h]
\begin{center}
\includegraphics[scale=0.8]{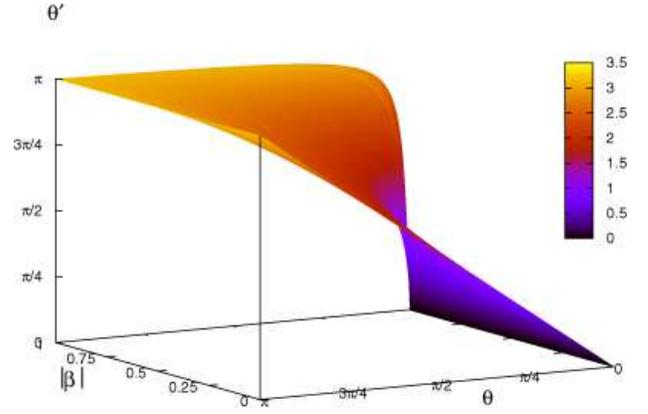}
\end{center}
\caption{Comportamiento del efecto de aberraci\'on con respecto a la norma del boost
($\mid\boldsymbol{\mathbf{\protect\beta}}\mid$) y $\protect\theta$ el
\'angulo entre la direcci\'on de propagaci\'on $\boldsymbol{\mathbf{k}}$ y
el boost $\boldsymbol{\mathbf{\protect\beta}}$.}
\label{base}
\end{figure}

El t\'{e}rmino de aberraci\'{o}n [ecuaci\'{o}n \eqref{a2}] viene dado por
\begin{equation}
\boldsymbol{{\mathbf{k^{\prime }}}}=\frac{\nu }{\nu ^{\prime }}\left\{
\boldsymbol{{\mathbf{k}}}-\gamma \boldsymbol{{\mathbf{\beta }}}\left( 1-%
\frac{\gamma \boldsymbol{{\mathbf{\beta }}}\centerdot \boldsymbol{{\mathbf{k}%
}}}{\gamma +1}\right) \right\} ,  \label{apendice3}
\end{equation}%
donde el t\'{e}rmino $\boldsymbol{{\mathbf{k}}}$ indica la direcci\'{o}n de
propagaci\'{o}n de la radiaci\'{o}n. Debido a que en la ecuaci\'{o}n %
\eqref{apendice3} se presenta el producto escalar $\boldsymbol{{\mathbf{%
\beta }}}\centerdot \boldsymbol{{\mathbf{k}}}$, el t\'{e}rmino de aberraci%
\'{o}n depende de las variables independientes $\beta \equiv \mid
\boldsymbol{\mathbf{\beta }}\mid $ y $\theta $, siendo $\theta $ el \'{a}%
ngulo entre la direcci\'{o}n de propagaci\'{o}n $\boldsymbol{\mathbf{k}}$ y
el boost $\boldsymbol{\mathbf{\beta }}$, de manera que la expresi\'{o}n %
\eqref{apendice3} se puede escribir como
\begin{equation}
\boldsymbol{{\mathbf{k^{\prime }}}}=\frac{\nu }{\nu ^{\prime }}\left\{
\boldsymbol{{\mathbf{k}}}-\gamma \boldsymbol{{\mathbf{\beta }}}\left( 1-%
\frac{\gamma \beta \cos (\theta )}{\gamma +1}\right) \right\} .
\label{apendice4}
\end{equation}%
Como el efecto relativista de aberraci\'{o}n ecuaci\'{o}n \eqref{apendice4},
es realmente un efecto relativista sobre el \'{a}ngulo entre la direcci\'{o}%
n de propagaci\'{o}n $\boldsymbol{\mathbf{k}}$ y el boost $\boldsymbol{%
\mathbf{\beta }}$ ya que la norma del vector direcci\'{o}n de propagaci\'{o}%
n (ya sea $\mid \boldsymbol{\mathbf{k}}\mid $ como $\mid \boldsymbol{\mathbf{%
k^{\prime }}}\mid $) es siempre igual a la unidad \cite{confusion}, al
realizar el producto punto a ambos lados de la igualdad de la ecuaci\'{o}n %
\eqref{apendice4} con $\boldsymbol{\mathbf{\beta }}$ se obtiene que
\begin{equation}
\boldsymbol{\mathbf{\beta }}\centerdot \boldsymbol{{\mathbf{k^{\prime }}}}=%
\frac{\nu }{\nu ^{\prime }}\boldsymbol{\mathbf{\beta }}\centerdot \left\{
\boldsymbol{{\mathbf{k}}}-\gamma \boldsymbol{{\mathbf{\beta }}}\left( 1-%
\frac{\gamma \beta \cos (\theta )}{\gamma +1}\right) \right\},
\label{apendice5}
\end{equation}%
definiendo a $\theta ^{\prime }$ el \'{a}ngulo entre $\boldsymbol{\mathbf{k}%
^{\prime }}$ y el boost, la relaci\'{o}n \eqref{apendice5} queda de la forma
\begin{equation}
\cos (\theta ^{\prime })=\frac{\nu }{\nu ^{\prime }}\left\{ \cos (\theta
^{)}-\gamma \beta \left( 1-\frac{\gamma \beta \cos (\theta )}{\gamma +1}%
\right) \right\},
\label{apendice6}
\end{equation}%
y reemplazando el factor de corrimiento \eqref{apendicea2} en la relaci\'{o}n \eqref{apendice6}, se encuentra que el efecto de aberraci\'{o}n es
\begin{equation}
\cos (\theta ^{\prime})=\frac{\cos (\theta )-\beta }{1-\beta \cos (\theta )},
\label{apendice7}
\end{equation}
donde se ha utilizado que $\gamma ^{2}\beta ^{2}=$ $\gamma ^{2}-1$. El comportamiento del efecto de aberraci\'on
\eqref{apendice7} se puede observar en la figura \ref{base}.

\begin{figure}[!h]
\begin{center}
\includegraphics[scale=0.8]{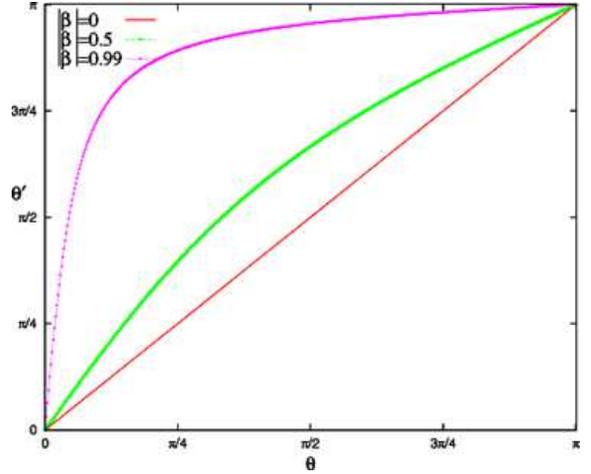}
\end{center}
\caption{Comportamiento del efecto de aberraci\'on a boost constantes}
\label{boost-ctes}
\end{figure}

\begin{figure}[!h]
\begin{center}
\includegraphics[scale=0.8]{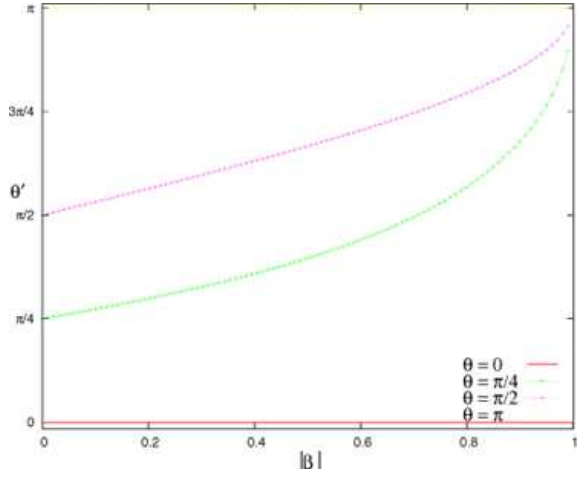}
\end{center}
\caption{Comportamiento del efecto de aberraci\'on a \'angulos constantes}
\label{angles-ctes}
\end{figure}

Una visi\'on detallada del comportamiento del efecto de aberraci\'on (figura %
\ref{base}) se puede observar en la figura \ref{boost-ctes} (boost constantes en magnitud) y en la figura \ref{angles-ctes}
(\'angulos constantes).

En la figura \ref{boost-ctes} se observa como el efecto de aberraci\'on para un boost nulo es igual a la unidad, ya
que esto significa que son dos sistemas que est\'an en reposo relativo entre si, pero que sus ejes pueden estar rotando un
\'angulo $\theta$ entre ellos. Para un boost diferente de cero (e.g., $\beta=0.5$ \'o $\beta=0.99$) se puede observar como
el efecto de aberraci\'on no solo va a depender de la norma de la velocidad relativa entre los sistemas inerciales sino de
su direcci\'on de manera que la aberraci\'on $\theta^{\prime}$ crece hacia su valor l\'{\i}imite a medida que la norma del
boost y su direcci\'on (i.e., $\beta$ y $\theta$, respectivamente) tienden hacia sus valores extremos, respectivamente.

En la figura \ref{angles-ctes} se observa como el efecto de aberraci\'on para un \'angulo $\theta$ igual a $0$ \'o $\pi$
es independiente de la velocidad relativa entre los marcos de referencia (i.e., el boost), mientras que para \'angulos con valor
diferente a los anteriores valores (e.g., $\theta=\pi/2$ \'o $\theta=\pi/4$) la aberraci\'on $\theta^{\prime}$ crece hacia
su valor l\'{\i}imite a medida que la norma del boost y su direcci\'on (i.e., $\beta$ y $\theta$, respectivamente) tienden hacia sus valores extremos, respectivamente.
\section{M\'etodo Directo}\label{referi}
De acuerdo con \eqref{eqreferi}, la ecuaci\'on de transferencia radiativa se puede escribir
como:
\begin{equation}
k^{\mu}\partial_{\mu}I=g.
\end{equation}
Dada la invarianza de $I$ y $g$, una manera compacta de demostrar la
invarianza de la ecuaci\'on de transferencia radiativa es mostrar que el
operador
\begin{equation}
k^{\mu}\partial_{\mu},
\end{equation}
es invariante. Para ello se consideran las reglas de transformaci\'on
\begin{equation}
k^{\mu}=\Lambda^{\mu}_{{\rho}'}k^{{\rho}'},
\end{equation}
\begin{equation}
\partial_{\mu}=\Lambda^{{\sigma}'}_{\mu}\partial_{{\sigma}'},
\end{equation}
donde $\Lambda$ es una transformaci\'on ortogonal con \'algebra 
\begin{equation}
\Lambda^{\mu}_{{\sigma}'}\Lambda^{{\rho}'}_{\mu}=\delta^{{\rho}'}_{{\sigma}'}.
\end{equation}
La transformaci\'on $\Lambda$ contiene como caso particular los \emph{boost} de
Lorentz. De est\'a manera es directo mostrar que
\begin{equation}
k^{\mu}\partial_{\mu}=k^{{\rho}'}\partial_{{\rho}'},
\end{equation}
con lo cual se muestra de una manera directa la invarianza de la ecuaci\'on de
transferencia radiativa. Sin embargo, dado el car\'acter educativo que
pretende el art\'{\i}culo, nuestro enfoque se realza m\'as sobre los efectos
cin\'ematicos y su interpretaci\'on f\'{\i}sica, dado que en los libros de
texto y en la bibliograf\'{\i}a especializada se dejan de lado y no se
explotan a fondo sus riquezas.
\bibliography{paper-rt-v2}
\end{document}